\documentclass[11pt]{article}
\usepackage[english,activeacute]{babel}
\usepackage{epsfig}
\usepackage{amsmath}
\usepackage{epic}
\usepackage{amssymb}
\usepackage{fancybox}
\usepackage{wrapfig}
\usepackage[font=footnotesize,labelsep=period]{caption}
\usepackage{amsthm}
\usepackage{float}
 \usepackage[linktocpage=true,colorlinks]{hyperref}
 \usepackage[capitalise]{cleveref}
\usepackage{xcolor}
\definecolor{dark-red}{rgb}{0.6,0.15,0.15}
\definecolor{dark-blue}{rgb}{0.15,0.15,0.8}
\definecolor{medium-blue}{rgb}{0,0,0.6}
\hypersetup{
    colorlinks, linkcolor={dark-red},
    citecolor={dark-blue}, urlcolor={medium-blue}
}
\allowdisplaybreaks
\usepackage{booktabs}

\usepackage{wrapfig}
\usepackage{cleveref}

\newcommand{\dashedrightarrow}[2][]{\ext@arrow 0359\rightarrowfill@@{#1}{#2}}

\usepackage[all]{xy}
\usepackage{xypic, color, graphics}

\theoremstyle{plain}

\theoremstyle{definition}

 \def\v #1{\vert #1\vert}             
 
 \def\m #1 #2{(-1)^{{\v #1} {\v #2}}} 

 \let \m=\medskip

\def\bea{\begin{eqnarray}}
 \def\eea{\end{eqnarray}}
\newcommand{\beq}{\begin{eqnarray}}
\newcommand{\eeq}{\end{eqnarray}}
\newcommand{\ba}{\begin{array}}
\newcommand{\ea}{\end{array}}
\newcommand{\be}{\begin {equation}}
\newcommand{\ee}{\end{equation}}
\def\picture #1 by #2 (#3){
  \vbox to #2{
    \hrule width #1 height 0pt depth 0pt
    \vfill
    \special{picture #3} 
    }
  }
\def\scaledpicture #1 by #2 (#3 scaled #4){{
  \dimen0=#1 \dimen1=#2
  \divide\dimen0 by 1000 \multiply\dimen0 by #4
  \divide\dimen1 by 1000 \multiply\dimen1 by #4
  \picture \dimen0 by \dimen1 (#3 scaled #4)}
  }
  
\usepackage{fancyhdr}

\usepackage[Bjornstrup]{fncychap}
\usepackage{fancyhdr}

\usepackage{appendix}
\usepackage[nottoc,numbib]{tocbibind}
\begin{document}

\centerline{\Large \bf Nonisospectral $1+1$ hierarchies arising from}\vskip 0.25cm
\centerline{\Large \bf a Camassa--Holm hierarchy in $2+1$ dimensions.} \vskip 0.25cm

\centerline{P.G. Est\'evez and C.
Sard\'on}
\vskip 0.5cm
\centerline{Department of Fundamental Physics, University of
Salamanca,}
\centerline{Plza. de la Merced s/n, 37.008, Salamanca, Spain.}

\vskip 1cm

\begin{abstract}
The non-isospectral problem (Lax pair) associated with a hierarchy in 2+1 dimensions that generalizes the
well known Camassa-Holm hierarchy is presented. Here, we have investigated the non-classical Lie symmetries
of this Lax pair when the spectral parameter is considered as a field. These symmetries can be written in
terms of five arbitrary constants and three arbitrary functions. Different similarity reductions associated
with these symmetries have been derived. Of particular interest are the reduced hierarchies whose $1+1$ Lax
pair is also non-isospectral.
\end{abstract}

\section{Introduction}

The identification of the Lie symmetries of a given partial differential equation (PDE) is  an instrument of
primary importance in order to solve such an equation \cite{stephani}. A standard method for finding
solutions of PDEs is that of reduction using Lie symmetries: each Lie symmetry allows a reduction of the PDE
to a new equation with the number of independent variables reduced by one \cite{bluco}, \cite{olver}. To a
certain extent this procedure gives rise to the ARS conjecture \cite{ars}, which establishes that a PDE is
integrable in the sense of Painlev\'e  \cite {pp} if all its reductions pass the Painlev\'e test
\cite{Weiss}. This means that the solutions of a PDE can be achieved by solving its reductions to ordinary
differential equations (ODE). Classical \cite{stephani} and non-classical \cite{bluco}, \cite{olver} Lie
symmetries are the usual way for identifying the reductions.

\section{The 2+1 Camassa-Holm Hierarchy}
\subsection*{Lax pair}
A generalization to $2+1$ dimensions of the celebrated Camassa-Holm  hierarchy (henceforth CHn2+1) was
presented in \cite{ep05-1}. By using reciprocal transformations,  this hierarchy was proved to be equivalent
to $n$ copies of the AKNS equation in 2+1 variables  \cite{ep04}, \cite{kp98}. It is well known that the 2+1
AKNS equation has the Painlev\'e property, and its non-isospectral Lax pair can be obtained by means of the
singular manifold method \cite{ep04}. Therefore we can use the inverse reciprocal transformation to obtain
the Lax pair of CHn2+1 \cite{ep05-1}. This Lax pair is also  a non-isospectral one that can be written in
terms of $n+1$ fields as follows:
\begin{eqnarray}&&\psi_{xx}-\left(\frac{1}{4}-\frac{\lambda}{2}M\right)\psi=0\nonumber\\
&&\psi_y-\lambda^n\psi_t+{\cal \hat A} \psi_x-\frac{{\cal \hat A}_x}{2}\psi=0,\label{2.1}
\end{eqnarray}
where
\begin{equation}{\cal \hat A}=\sum_{j=1}^n \lambda^{(n-j+1)}U^{[j]} \label{2.2}\end{equation} and
$$M=M(x,y,t),\quad U^{[j]}=U^{[j]}(x,y,t), \,j=1...n.$$
\subsection*{Non-isospectrality and equations}
The compatibility condition between equations (\ref{2.1}) yields the non-isospectral condition
\begin{equation}\lambda_y-\lambda^n\lambda_t=0,\quad\quad\quad  \lambda_x=0,\label{2.3}
\end{equation}
as well as the equations
\begin{eqnarray}&&M_y=U^{[n]}_{x}-U^{[n]}_{xxx} \nonumber\\
&&M_t= U^{[1]}M_x+2MU^{[1]}_x\\
&&U^{[j]}M_x+2MU^{[j]}_x=U^{[j-1]}_{x}-U^{[j-1]}_{xxx},\quad j=2..n.\nonumber \label{2.4}
\end{eqnarray}

\subsection*{Recursion operator and hierarchy}
The above equations can be written in more compact form by defining the operators:
\begin{equation}J=\frac{\partial }{\partial x}-\frac{\partial }{\partial x^3},
 \quad\quad\quad K=M\frac{\partial }{\partial x}+\frac{\partial }{\partial x}M. \label{2.5}\end{equation}
Equations (2.4) are therefore:
\begin{eqnarray}&&M_y=JU^{[n]} \nonumber\\
&&M_t= KU^{[1]} \label{2.6}\\
&&KU^{[j]}=JU^{[j-1]},\quad j=2..n,\nonumber
\end{eqnarray}
which yields the hierarchy:
\begin{equation}M_y=R^nM_t \label{2.7}\end{equation}
where the recursion operator is:
\begin{equation}R= JK^{-1}.\label{2.8}\end{equation}
Solutions of these equations were studied in \cite{ep05-2}. The positive and negative, \cite{AF88},
\cite{cgp97}, $1+1$ Camassa-Holm hierarchies can be obtained by setting $\frac{\partial }{\partial
y}=\frac{\partial }{\partial x}$ or $\frac{\partial }{\partial y}=\frac{\partial }{\partial t}$ respectively
\cite{ep05-1}.

The $n=-1$ case of (\ref{2.4}) has been considered in \cite{cgp97} and \cite{ivanov}. There are also
different generalizations of the Camassa-Holm hierarchy to $2+1$ dimensions arising from different (although
isospectral)  spectral problems \cite{kraenkel}, \cite{ortenzi}.
\section{Non classical symmetries of the CHn2+1 spectral problem}
\subsection*{Lie point symmetries}
Here, we are interested  in the Lie symmetries of the Lax pair (\ref{2.1}). Naturally, the symmetries of
equations (2.4)  are interesting in themselves, but we  also wish to know
 how \textbf{the eigenfunction and the spectral parameter transform under the action of a Lie symmetry}.
  More precisely, we wish to know what these fields look like under the reduction associated with each symmetry.
This is why we shall proceed to write the infinitesimal Lie point transformation of the variables
   and fields that appear in the spectral problem (\ref{2.1}).
   We have proved the benefits of such a procedure \cite{leg} in a previous paper \cite{egp05}.

   In the present case, it is important to note that the
   spectral parameter $\lambda(y,t)$ is not a constant, and therefore that it should be considered as an additional
   field satisfying (\ref{2.3}). This
    means that we are actually looking for the Lie point symmetries of  equations
   (\ref{2.1}) together with (\ref{2.3}).

   The infinitesimal form of the Lie point symmetry that we are considering is:

\begin{eqnarray}
 x' &=& x + \varepsilon\,\xi_1(x,y,t,\lambda,\psi,M,U^{[j]}) + O(\varepsilon^2)\nonumber\\
y' &=&y + \varepsilon\,\xi_2(x,y,t,\lambda,\psi,M,U^{[j]}) + O(\varepsilon^2)\nonumber\\
t'  &=& t + \varepsilon\,\xi_3(x,y,t,\lambda,\psi,M,U^{[j]}) + O(\varepsilon^2)\label{3.1}\\
\psi'  &=& \psi + \varepsilon\,\phi_1(x,y,t,\lambda,\psi,M,U^{[j]}) + O(\varepsilon^2)\nonumber
\\
\lambda'  &=& \lambda + \varepsilon\,\phi_2(x,y,t,\lambda,\psi,M,U^{[j]}) + O(\varepsilon^2)\nonumber
\\M'  &=& M+ \varepsilon\,\Theta_0(x,y,t,\lambda,\psi,M,U^{[j]}) + O(\varepsilon^2)\nonumber\\
(U^{[i]})' &=& U^{[i]}+ \varepsilon\,\Theta_j(x,y,t,\lambda,\psi,M,U^{[j]}) + O(\varepsilon^2),\quad
i,j=1..n. \nonumber
\end{eqnarray} where
$\epsilon$ is the group parameter. The associated Lie algebra of infinitesimal symmetries is the set of
vector fields of the form:
\begin{equation} X = \xi_1\frac {\partial}{\partial x} + \xi_2\frac {\partial}{\partial y}+\xi_3\frac
{\partial}{\partial t}+\phi_1\frac {\partial}{\partial \psi}+\phi_2\frac {\partial}{\partial
\lambda}+\Theta_0\frac {\partial}{\partial M}+\sum_{j=1}^n\Theta_j\frac {\partial}{\partial
U^{[j]}}.\label{3.2}\end{equation}

 We  also need to know
how the derivatives of the fields transform under the Lie symmetry. This means that we have  to introduce the
``prolongations" of the action of the group  to the different derivatives that appear in (\ref{2.1}) and
(\ref{2.3}). Exactly how to calculate the prolongations is a very well known procedure whose technical
details can be found in \cite{stephani}.

 It is therefore necessary  that the Lie  transformation should leave (\ref{2.1}) and (\ref{2.3}) invariant.
 This yields an overdetermined  system of equations for the
infinitesimals $\xi_1(x,y,t,\lambda,\psi,M,U^{[j]})$, $\xi_2(x,y,t,\lambda,\psi,M,U^{[j]})$, $
\xi_3(x,y,t,\lambda,\psi,M,U^{[j]})$, $\phi_1(x,y,t,\lambda,\psi,M,U^{[j]})$,
$\phi_2(x,y,t,\lambda,\psi,M,U^{[j]}),$ and $ \Theta_i(x,y,t,\lambda,\psi,M,U^{[j]})$.

This is the  \textbf{classical method} \cite{stephani} of finding Lie symmetries, and it can be summarized as
follows
\begin{itemize}

\item  Calculation of the prolongations of the derivatives of the fields that appear in (\ref{2.1}) and
(\ref{2.3}).
\item  Substitution of the transformed fields (\ref{3.1}) and their derivatives in (\ref{2.1}) and
(\ref{2.3}).
\item  Set  all the coefficients in $\epsilon$ at $0$.
\item Substitution of the prolongations.
\item  $\psi_{xx},\psi_y$ and $\lambda_y$ can be substituted by using (\ref{2.1}) and
(\ref{2.3}).
\item  The system of equations for the infinitesimals can be obtained by setting  each coefficient in
the different remaining derivatives of the fields at zero.
\end{itemize}

\subsection*{Non-classical symmetries}
There is a generalization of the classical method that determines the  \textbf{non-classical or conditional
symmetries} \cite{bluco}, \cite{olver}. In this case we are looking for symmetries that leave invariant not
only the equations but also the so called ``invariant  surfaces", which in our case are:
\begin{eqnarray}
 \phi_1&=& \xi_1 \psi_x+\xi_2\psi_y+\xi_3\psi_t\nonumber\\
 \phi_2&=& \xi_2\lambda_y+\xi_3\lambda_t\nonumber\\
 \Theta_0&=& \xi_1 M_x+\xi_2M_y+\xi_3M_t \label{3.3}\\
 \Theta_j&=& \xi_1 U^{[j]}_x+\xi_2U^{[j]}_y+\xi_3U^{[j]}_t,\quad j=1..n.\nonumber
\end{eqnarray}

These non-classical symmetries are the symmetries that we address below. The method for calculating these
symmetries is the same as the one we have described for the classical ones complemented with equations
(\ref{3.3}), that must also be combined with step 4 to eliminate as many derivatives of the fields as
possible, depending on whether all of the $\xi_i$ are different from zero or not. This is why we have  to
distinguish three different types of non-classical symmetries.
\begin{itemize}
\item $\xi_3=1.$
\item $\xi_3=0,\,\xi_2=1.$
\item $\xi_3=0,\,\xi_2=0,\, \xi_1=1.$
\end{itemize}
Note that owing to (\ref{3.3}), there is not restriction in selecting $\xi_j=1$ when $\xi_j\neq 0$
\cite{olver}. In the following sections we shall determine these three types of symmetries of the Lax pair
and its reduction to $1+1$ dimensions by solving the characteristic equation
\begin{equation}
\frac{dx}{\xi_1}=\frac{dy}{\xi_2}=\frac{dt}{\xi_3}=\frac{d\psi}{\phi_1}=\frac{d\lambda}{\phi_2}=\frac{dM}{\Theta_0}=\frac{dU^{[j]}}{\Theta_j}.\label{3.4}
\end{equation}
The advantage of our approach of working with the Lax pair instead of the equations of the hierarchy lies in
the fact that we can  obtain the reduced eigenfuntion and the reduced spectral parameter at the same time.
which as we shall see, in many cases is not a trivial matter. The equations of  the reduced hierarchies can
be explicitly  obtained from the reduced spectral problem and we shall write them in all the cases.

Of course the calculation of the symmetries is tedious , and we have used the MAPLE symbolic package  to
handle these calculations. For the benefit of the reader, we shall omit the technical details.

\section{Non-classical symmetries for $\xi_3=1$}
\subsection*{Calculation of symmetries}
In this case (\ref{3.3}) allows us to eliminate the derivatives with respect to $t$
\begin{eqnarray}
\psi_t&=& \phi_1-\xi_1 \psi_x-\xi_2\psi_y\nonumber\\
\lambda_t&=&  \phi_2-\xi_2\lambda_y\nonumber\\
 M_t&=& \Theta_0-\xi_1 M_x-\xi_2M_y \label{4.1}\\
U^{[j]}_t &=&\Theta_j- \xi_1 U^{[j]}_x-\xi_2U^{[j]}_y,\quad j=1..n.\nonumber
\end{eqnarray}
If we add (\ref{4.1}) to the five steps listed above for the calculation of non-classical symmetries, we
obtain (after long but straightforward calculations)  the following symmetries:
\begin{eqnarray}
 \xi_1&=& \frac{S_1}{S_3}\nonumber\\
 \xi_2&=& \frac{S_2}{S_3}\nonumber\\
 \xi_3&=& 1\nonumber\\
  \phi_1&=& \frac{1}{S_3}\left(\frac{1}{2}\frac{\partial S_1}{\partial x}+a_0\right)\psi\label{4.2}\\
   \phi_2&=& \frac{1}{S_3}\left(\frac{a_3-a_2}{n}\right)\lambda\nonumber\\
 \Theta_0&=& \frac{1}{S_3}\left(-2\frac{\partial S_1}{\partial x}+\frac{a_2-a_3}{n}\right)M\nonumber\\
  \Theta_1&=& \frac{1}{S_3}\left(U^{[1]}\left(\frac{\partial S_1}{\partial x}-a_3\right)-\frac{\partial S_1}{\partial t}\right)\nonumber\\
 \Theta_j&=& \frac{1}{S_3}\left(\frac{\partial S_1}{\partial x}-a_2\frac{j-1}{n}-a_3\frac{n-j+1}{n}\right)U^{[j]},\quad j=2..n,\nonumber
\end{eqnarray}
where
\begin{eqnarray}
 S_1&=& S_1(x,t)= A_1(t)+B_1(t)\,e^x+C_1(t)\,e^{-x},\nonumber\\
S_2&=& S_2(y)=a_2y+b_2,\label{4.3}\\
S_3&=& S_3(t)= a_3t+b_3.\nonumber
\end{eqnarray}
$A_1(t),B_1(t),C_1(t)$ are arbitrary functions of $t$. Furthermore, $a_0, a_2, b_2, a_3, b_3$ are arbitrary
constants, such that $a_3$ and $b_3$ cannot  at the same time be $0$.
\subsection*{Classification of the reductions}We have, therefore, several different reductions depending on which arbitrary functions and/or constants are or are not zero.
We shall use the following classification
\begin{itemize}
\item Type I: Corresponding to selecting $A_1(t)\neq 0$, $B_1(t)=C_1(t)=0$.
\item Type II: Corresponding to selecting $B_1(t)\neq 0$, $A_1(t)=C_1(t)=0$. As we shall show in Appendix I,
this case yields  the same reduced spectral problems as those obtained for Type I, although the reductions
are different.
\item Type III: Corresponding to selecting $C_1(t)\neq 0$, $A_1(t)=B_1(t)=0$. It is easy to see that this case
is equivalent to II owing to the invariance of the Lax pair under the transformation $x\rightarrow -x$,
$y\rightarrow -y$, $t\rightarrow -t$. Below we only consider cases I and II.
\end{itemize}
In each of the cases listed before we have different subcases, depending on the values of the constants $a_j$
and $b_j$.  We have the following 5 independent  possibilities:

\begin{itemize}

\item Case 1:  $a_2=0,\quad a_3=0$; $\quad\quad b_2= 0$.
\item Case 2:  $a_2=0,\quad a_3=0$;$\quad\quad\, b_2\neq 0$.
\item Case 3:  $a_2=0,\quad a_3\neq 0$; $\quad\quad b_2= 0$.
\item Case 4:  $a_2=0,\quad a_3\neq 0$;$\quad\quad\, b_2\neq 0$.
\item Case 5:  $a_2\neq 0$;

\end{itemize}
We can obtain 5 different non-trivial reductions: I.i $i=1..5$. We shall see each reduction separately by
obtaining the reduced variables, the reduced fields, the transformation of the spectral parameter and the
eigenfuction and, finally, the reduced spectral problem and the corresponding reduced hierarchy. Furthermore,
there are several interesting reductions, especially those that also have a non-isospectral parameter in $1+1$
dimensions . Let us summarize the results:

\subsection*{I.1) $B_1(t)=C_1(t)=0,\,A_1(t)\neq 0,\quad a_2=0,\, a_3=0,\, b_2= 0.$}
By solving the characteristic equation (\ref{3.4}), we have the following results
\begin{itemize}
\item Reduced variables: $\quad z_1=x-\frac{1}{b_3}\int A_1(t)\,dt $, $\quad z_2=y $
\item Spectral parameter: $\quad \lambda(y,t)=\lambda_0$
\item Reduced Fields:
\begin{eqnarray} &&\psi(x,y,t)={\displaystyle e}^{{\displaystyle  \left(\frac{a_0t}{b_3}\right)}}\,{\displaystyle e}^
{{\displaystyle\left( \frac{\lambda_0^n
a_0z_2}{b_3}\right)}}\,\Phi(z_1,z_2)\nonumber\\
&&M(x,y,t)=H(z_1,z_2)\nonumber\\
 &&U^{[1]}(x,y,t)=V^{[1]}(z_1,z_2)-\frac{A_1}{b_3}\nonumber\\ &&U^{[j]}(x,y,t)=V^{[j]}(z_1,z_2)\nonumber\end{eqnarray}

\item Reduced Spectral problem:
\begin{eqnarray}&&\Phi_{z_1z_1}-\left(\frac{1}{4}-\frac{\lambda_0}{2}H\right)\Phi=0\nonumber\\
&&\Phi_{z_2}+{\cal \hat B}\Phi_{z_1}-\frac{{\cal \hat B}_{z_1}}{2}\Phi=0\label{4.4}
\end{eqnarray}
where
\begin{equation} {\cal \hat B}=\sum_{j=1}^n \lambda_0^{(n-j+1)}V^{[j]}(z_1,z_2) \label{4.5} \end{equation}

\item Reduced Hierarchy: The compatibility condition of (\ref{4.4}) yields:
\begin{eqnarray}&&\frac{\partial^3 V^{[n]}}{\partial z_1^3}-\frac{\partial V^{[n]}}{\partial z_1}+\frac{\partial H}{\partial z_2}=0\nonumber\\
&&2H\frac{\partial V^{[1]}}{\partial z_1}+V^{[1]}\frac{\partial H}{\partial z_1}=0\\
&&2H\frac{\partial V^{[j+1]}}{\partial z_1}+V^{[j+ 1]}\frac{\partial H}{\partial z_1}+\frac{\partial^3
V^{[j]}}{\partial z_1^3}-\frac{\partial V^{[j]}}{\partial z_1}=0.\nonumber \label{4.6}
\end{eqnarray}
which is  \textbf{the positive Camassa-Holm hierarchy}, whose first component ($n=1$) is a modified Dym
equation \cite{AF88}, \cite{ep05-1}.
\end{itemize}

\subsection*{I.2) $B_1(t)=C_1(t)=0,\,A_1(t)\neq 0,\quad a_2=0,\, a_3=0,\, b_2\neq 0.$}
By solving the characteristic equation (\ref{3.4}), we have the following results
\begin{itemize}
\item Reduced variables: $\quad z_1=x-\frac{1}{b_3}\int A_1(t)\,dt$, $\quad z_2=\frac{y}{b_2}-\frac{t}{b_3} $
\item Spectral parameter: $\quad \lambda(y,t)=\left(\frac{b_3}{b_2}\right)^{\left(\frac{1}{n}\right)}\lambda_0$
\item Reduced Fields:  \begin{eqnarray} && \psi(x,y,t)=
{\displaystyle e}^{\left({\displaystyle \frac{a_0t}{b_3}}\right)}\,{\displaystyle e}
^{{\displaystyle\left(\frac{\lambda_0^n a_0z_2}{1+\lambda_0^n
}\right)}}\,\Phi(z_1,z_2)\nonumber \\
&& M(x,y,t)=\left(\frac{b_2}{b_3}\right)^{\left(\frac{1}{n}\right)}\,H(z_1,z_2)\nonumber\\
&& U^{[1]}(x,y,t)=\left(\frac{1}{b_3}\right)V^{[1]}(z_1,z_2)-\frac{A_1}{b_3}\nonumber\\
&& U^{[j]}(x,y,t)=\left(\frac{1}{b_3}\right)\left(\frac{b_2}{b_3}\right)^{\left(
\frac{1-j}{n}\right)}V^{[j]}(z_1,z_2)\nonumber\end{eqnarray}

\item Reduced Spectral problem:
\begin{eqnarray}&&\Phi_{z_1z_1}-\left(\frac{1}{4}-\frac{\lambda_0}{2}H\right)\Phi=0\nonumber\\
&&\Phi_{z_2}\left(1+\lambda_0^n\right)+{\cal \hat B}\Phi_{z_1}-\frac{{\cal \hat
B}_{z_1}}{2}\Phi=0\label{4.7}\end{eqnarray} where
\begin{equation} {\cal \hat B}=\sum_{j=1}^n \lambda_0^{(n-j+1)}V^{[j]}(z_1,z_2)  \label{4.8}\end{equation}

\item Reduced Hierarchy: The compatibility condition of (\ref{4.7}) yields:
\begin{eqnarray}&&\frac{\partial^3 V^{[n]}}{\partial z_1^3}-\frac{\partial V^{[n]}}{\partial z_1}+\frac{\partial H}{\partial z_2}=0\nonumber\\
&&2H\frac{\partial V^{[1]}}{\partial z_1}+V^{[1]}\frac{\partial H}{\partial z_1}+\frac{\partial H}{\partial z_2}=0\\
&&2H\frac{\partial V^{[j+1]}}{\partial z_1}+V^{[j+ 1]}\frac{\partial H}{\partial z_1}+\frac{\partial^3
V^{[j]}}{\partial z_1^3}-\frac{\partial V^{[j]}}{\partial z_1}=0.\nonumber \label{4.9}
\end{eqnarray}

\end{itemize}

\subsection*{I.3) $B_1(t)=C_1(t)=0,\,A_1(t)\neq 0,\quad a_2=0,\, a_3\neq 0,\, b_2=0.$}
By solving the characteristic equation (\ref{3.4}), we have the following results
\begin{itemize}
\item Reduced variables: $\quad z_1=x-\int \frac{A_1(t)}{S_3(t)} \,dt$, $\quad z_2=a_3y $
\item Spectral parameter: In this case the reduction of the spectral parameter is a non-trivial one that yields
 $$\quad \lambda(y,t)=S_3^{\left(\frac{1}{n}\right)}\Lambda(z_2)$$
 where $\Lambda(z_2)$ is the reduced spectral parameter.

\item Reduced Fields:
\begin{eqnarray} &&  \psi(x,y,t)=\Lambda(z_2)^{{\displaystyle \left(\frac{a_0 n}{a_3}\right)}}
S_3^{{\displaystyle\left(\frac{a_0 }{a_3}\right)}}\Phi(z_1,z_2)\nonumber\\
&& M(x,y,t)=S_3^{\left(-\frac{1 }{n}\right)}H(z_1,z_2)\nonumber\\
 && U^{[1]}(x,y,t)=\frac{a_3}{S_3}V^{[1]}(z_1,z_2)-\frac{A_1}{S_3}\nonumber\\
&& U^{[j]}(x,y,t)=a_3\,S_3^{\left(\frac{j-1 }{n}-1\right)}V^{[j]}(z_1,z_2)\nonumber\end{eqnarray}

\item Reduced Spectral problem:
\begin{eqnarray}&&\Phi_{z_1z_1}-\left(\frac{1}{4}-\frac{\Lambda(z_2)}{2}H\right)\Phi=0\nonumber\\
&&\Phi_{z_2}+{\cal \hat B}\Phi_{z_1}-\frac{{\cal \hat B}_{z_1}}{2}\Phi=0\label{4.10}
\end{eqnarray}
where
\begin{equation} {\cal \hat B}=\sum_{j=1}^n \Lambda(z_2)^{(n-j+1)}V^{[j]}(z_1,z_2)  \label{4.11}\end{equation}
and $\Lambda(z_2)$ satisfies \textbf{the non-isospectral condition}
 \begin{equation} n\frac {d\Lambda(z_2)}{d\,z_2}-\Lambda(z_2)^{(n+1)}=0\label{4.12}\end{equation}
\item Reduced Hierarchy: The compatibility condition of (\ref{4.10}) yields \textbf{the autonomous hierarchy}:
\begin{eqnarray}&&\frac{\partial^3 V^{[n]}}{\partial z_1^3}-
\frac{\partial V^{[n]}}{\partial z_1}+\frac{\partial H}{\partial z_2}=0\nonumber\\
&&2H\frac{\partial V^{[1]}}{\partial z_1}+V^{[1]}\frac{\partial H}{\partial z_1}+\frac{H}{n}=0\\
&&2H\frac{\partial V^{[j+1]}}{\partial z_1}+V^{[j+ 1]}\frac{\partial H}{\partial z_1}+\frac{\partial^3
V^{[j]}}{\partial z_1^3}-\frac{\partial V^{[j]}}{\partial z_1}=0.\nonumber \label{4.13}
\end{eqnarray}
\end{itemize}

\subsection*{I.4) $B_1(t)=C_1(t)=0,\,A_1(t)\neq 0,\quad a_2=0,\, a_3\neq 0,\, b_2\neq 0.$}
By solving the characteristic equation (\ref{3.4}), we have the following results
\begin{itemize}
\item Reduced variables: $\quad z_1=x-\int \frac{A_1(t)}{S_3(t)} \,dt$, $\quad z_2=\frac{a_3y}{b_2} -\textrm{ln}(S_3)$
\item Spectral parameter: In this case the reduction of the spectral parameter  yields
 $$\quad \lambda(y,t)=\left(\frac{S_3}{b_2}\right)^{\left(\frac{1}{n}\right)}\Lambda(z_2)$$
 where $\Lambda(z_2)$ is the reduced spectral parameter:

    \item Reduced Fields:
\begin{eqnarray} && \psi(x,y,t)=\Lambda(z_2)^{{\displaystyle\left(\frac{na_0}{a_3}\right)}}\,S_3^{{\displaystyle \left(\frac{a_0}{a_3}\right)}}\,\Phi(z_1,z_2)\nonumber \\
&&M(x,y,t)=\left(\frac{b_2}{S_3}\right)^{\left(\frac{1}{n}\right)}H(z_1,z_2)\nonumber\\
&& U^{[1]}(x,y,t)=\frac{a_3}{S_3}V^{[1]}(z_1,z_2)-\frac{A_1}{S_3}\nonumber \\
 &&U^{[j]}(x,y,t)=\frac{a_3}{S_3}\left(\frac{S_3}{b_2}\right)^{\left(\frac{j-1}{n}\right)}V^{[j]}(z_1,z_2)\nonumber\end{eqnarray}

\item Reduced Spectral problem:
\begin{eqnarray}&&\Phi_{z_1z_1}-\left(\frac{1}{4}-\frac{\Lambda(z_2)}{2}H\right)\Phi=0\nonumber\\
&&\Phi_{z_2}\left(1+\Lambda(z_2)^n\right)+{\cal \hat B}\Phi_{z_1}-\frac{{\cal \hat
B}_{z_1}}{2}\Phi=0\label{4.14}
\end{eqnarray}
where
\begin{equation} {\cal \hat B}=\sum_{j=1}^n \left(\Lambda(z_2)\right)^{(n-j+1)}V^{[j]}(z_1,z_2)  \label{4.15}\end{equation}
and $\Lambda(z_2)$ satisfies \textbf{the non-isospectral condition}
 \begin{equation} n\left(1+\Lambda(z_2)^n\right)\frac {d\Lambda}{d\,z_2}-\Lambda(z_2)^{n+1}=0\label{4.16}\end{equation}
\item Reduced Hierarchy: The compatibility condition of (\ref{4.14}) yields:
\begin{eqnarray}&&\frac{\partial^3 V^{[n]}}{\partial z_1^3}-
\frac{\partial V^{[n]}}{\partial z_1}+\frac{\partial H}{\partial z_2}=0\nonumber\\
&&2H\frac{\partial V^{[1]}}{\partial z_1}+V^{[1]}\frac{\partial H}{\partial z_1}+\frac{H}{n}+\frac{\partial H}{\partial z_2}=0\\
&&2H\frac{\partial V^{[j+1]}}{\partial z_1}+V^{[j+ 1]}\frac{\partial H}{\partial z_1}+\frac{\partial^3
V^{[j]}}{\partial z_1^3}-\frac{\partial V^{[j]}}{\partial z_1}=0.\nonumber \label{4.17}
\end{eqnarray}
Therefore, \textbf{although the Lax pair is non-isospectral, the reduced hierarchy is autonomous}
\end{itemize}

\subsection*{I.5) $B_1(t)=C_1(t)=0,\,A_1(t)\neq 0,\quad a_2\neq 0$}
By solving the characteristic equation (\ref{3.4}), we have the following results :
\begin{itemize}
\item Reduced variables: $\quad z_1=x-\int \frac{A_1(t)}{S_3(t)} \,dt$, $\quad z_2=S_2\,S_3^{{\displaystyle \left(-\frac{a_2}{a_3}\right)}}$
\item Spectral parameter: In this case the reduction of the spectral parameter  yields
 $$\quad \lambda(y,t)=S_3^{{\displaystyle \left(\frac{a_3-a_2}{a_3n}\right)}}\Lambda(z_2)$$
 where $\Lambda(z_2)$ is the reduced spectral parameter:

\item Reduced Fields:
\begin{eqnarray} &&  \psi(x,y,t)=\Lambda(z_2)^{{\displaystyle \left(\frac{na_0}{a_3-a_2}\right)}}S_3^{{\displaystyle \left(\frac{a_0}{a_3}\right)}}\,\Phi(z_1,z_2)\nonumber \\
&& M(x,y,t)=S_3^{{\displaystyle \left(\frac{a_2-a_3}{a_3n}\right)}}\,H(z_1,z_2)\nonumber\\
 &&U^{[1]}(x,y,t)=\left(\frac{a_2}{S_3}\right)V^{[1]}(z_1,z_2)-\frac{A_1}{S_3}\nonumber\\
&& U^{[j]}(x,y,t)=\left(\frac{a_2}{S_3}\right)S_3^{{\displaystyle \left(\frac{(a_3-a_2)(j-1)}{a_3n}\right)}}
 V^{[j]}(z_1,z_2)\nonumber\end{eqnarray}

\item Reduced Spectral problem:
\begin{eqnarray}&&\Phi_{z_1z_1}-\left(\frac{1}{4}-\frac{\Lambda(z_2)}{2}H\right)\Phi=0\nonumber\\
&&\Phi_{z_2}\left(1+z_2\Lambda(z_2)^n\right)+{\cal \hat B}\Phi_{z_1}-\frac{{\cal \hat
B}_{z_1}}{2}\Phi=0\label{4.18}
\end{eqnarray}
where
\begin{equation} {\cal \hat B}=\sum_{j=1}^n \Lambda(z_2)^{(n-j+1)}V^{[j]}(z_1,z_2)  \label{4.19}\end{equation}
and $\Lambda(z_2)$ satisfies \textbf{the non-isospectral condition}
 \begin{equation} n(1+z_2\Lambda(z_2)^{n})\frac {d\Lambda}{d\,z_2}-\frac{a_3-a_2}{a_2}\Lambda(z_2)^{(n+1)}=0\label{4.20}\end{equation}
\item Reduced Hierarchy: The compatibility condition of (\ref{4.18}) yields:
\begin{eqnarray}&&\frac{\partial^3 V^{[n]}}{\partial z_1^3}-
\frac{\partial V^{[n]}}{\partial z_1}+\frac{\partial H}{\partial z_2}=0\nonumber\\
&&2H\frac{\partial V^{[1]}}{\partial z_1}+V^{[1]}\frac{\partial H}{\partial z_1}+\frac{a_3-a_2}{a_2}\frac{H}{n}+z_2\frac{\partial H}{\partial z_2}=0\\
&&2H\frac{\partial V^{[j+1]}}{\partial z_1}+V^{[j+ 1]}\frac{\partial H}{\partial z_1}+\frac{\partial^3
V^{[j]}}{\partial z_1^3}-\frac{\partial V^{[j]}}{\partial z_1}=0.\nonumber \label{4.21}
\end{eqnarray}
\textbf{the Lax pair is non-isospectral and the reduced hierarchy is non-autonomous}
\item
Note that the singularity,  which apparently appears in the reductions when $a_3=0$,  can be easily removed by
considering that
$$\textrm{lim}_{a_3\rightarrow 0}\left(\frac{a_3t+b_3}{b_3}\right)^{(1/a_3)}=e^{t/b_3}$$
\end{itemize}

We refer  readers to Appendix I so that they can check that the spectral problems obtained in case II are not
different from those of case I
\section{Non classical symmetries for $\xi_3=0$, $\xi_2=1$}
\subsection{Calculation of the symmetries}
We can  now write
\begin{eqnarray}
\psi_y&=& \phi_1-\xi_1 \psi_x\nonumber\\
\lambda_y&=&  \phi_2\nonumber\\
 M_y&=& \Theta_0-\xi_1 M_x\label{5.1}\\
U^{[j]}_y &=&\Theta_j- \xi_1 U^{[j]}_x,\quad j=1..n\nonumber
\end{eqnarray}
 We can combine (\ref{5.1}) with  (\ref{2.3}) and  (2.4). This allows us to remove $\psi_{xx}, \psi_y,
 \psi_t,
 \lambda_y,$ $\lambda_t, M_y, U^{[j]}_y$ from the equation of the symmetries. In this case, we obtain the
 following
symmetries
\begin{eqnarray}
 \xi_1&=& \frac{S_1}{S_2}\nonumber\\
 \xi_2&=& 1\nonumber\\
 \xi_3&=& 0\nonumber\\
  \phi_1&=& \frac{1}{S_2}\left(\frac{1}{2}\frac{\partial S_1}{\partial x}+a_0\right)\psi\label{5.2}\\
   \phi_2&=& \frac{1}{S_2}\left(\frac{-a_2}{n}\right)\lambda\nonumber\\
 \Theta_0&=& \frac{1}{S_2}\left(-2\frac{\partial S_1}{\partial x}+\frac{a_2}{n}\right)M\nonumber\\
  \Theta_1&=& \frac{1}{S_2}\left(U^{[1]}\left(\frac{\partial S_1}{\partial x}\right)-\frac{\partial S_1}{\partial t}\right)\nonumber\\
 \Theta_j&=& \frac{1}{S_2}\left(\frac{\partial S_1}{\partial x}-a_2\frac{j-1}{n}\right)U^{[j]},\quad j=2..n.\nonumber
\end{eqnarray}
where $S_1$ and $S_2$ are those given in (\ref{4.3}).

Evidently we should consider that $a_2$ and $b_2$ cannot be $0$ at the same time.
\subsection*{Classification of the reductions}
In this case, one of the reduced variables is $t$. This means that the integrals that involve $S_1$ can be
performed without any restrictions for the functions $A_1(t),B_1(t), C_1(t)$. We have four different cases

\subsection*{IV.1: $a_2=0, \quad E=\sqrt{A_1^2-4B_1C_1}= 0$.}

\begin{itemize}
\item Reduced variables: $\quad z_1=\int\frac{dx}{S_1(x,t)}-\frac{y}{b_2}$, $z_2=\frac{t}{b_2} $

\item Spectral parameter:
 $\quad \lambda(y,t)=\lambda_0$

\item Reduced Fields:
\begin{eqnarray} && \psi(x,y,t)=\sqrt{S_1}\,{\displaystyle e}^{{\displaystyle \left(\frac{a_0y}{b_2}\right)}}
{\displaystyle e}^{{\displaystyle \left(\frac{a_0t}{b_2\lambda_0^n}\right)}}\,\Phi(z_1,z_2)\nonumber \\&&M(x,y,t)=\frac{H(z_1,z_2)}{S_1^2}\nonumber\\
 && U^{[1]}(x,y,t)=\frac{S_1}{b_2}\,V^{[1]}(z_1,z_2)+S_1\frac{d}{dt}\left(\int\frac{dx}{S_1(x,t)}\right)\nonumber\\
&& U^{[j]}(x,y,t)= \frac{S_1}{b_2}\,V^{[j]}(z_1,z_2)\nonumber\end{eqnarray}

\item Reduced Spectral problem:
\begin{eqnarray}&&\Phi_{z_1z_1}+\frac{\lambda_0}{2}H\Phi=0\nonumber\\
&&\lambda_0^n\Phi_{z_2}=({\cal \hat B}-1)\Phi_{z_1}-\frac{{\cal \hat B}_{z_1}}{2}\Phi\label{5.4}
\end{eqnarray}
where
\begin{equation} {\cal \hat B}=\sum_{j=1}^n \lambda_0^{(n-j+1)}V^{[j]}(z_1,z_2)  \label{5.5}\end{equation}

\item Reduced Hierarchy: The compatibility condition of (\ref{5.4}) yields \textbf{the autonomous hierarchy}:
\begin{eqnarray}&&\frac{\partial^3 V^{[n]}}{\partial z_1^3}-\frac{\partial H}{\partial z_1}=0\nonumber\\
&&2H\frac{\partial V^{[1]}}{\partial z_1}+V^{[1]}\frac{\partial H}{\partial z_1}-\frac{\partial H}{\partial z_2}=0\\
&&2H\frac{\partial V^{[j+1]}}{\partial z_1}+V^{[j+ 1]}\frac{\partial H}{\partial z_1}+\frac{\partial^3
V^{[j]}}{\partial z_1^3}=0.\nonumber \label{5.6}
\end{eqnarray}
\end{itemize}
\subsection*{IV.2: $a_2=0, \quad E=\sqrt{A_1^2-4B_1C_1}\neq 0$.}

\begin{itemize}
\item Reduced variables: $\quad z_1=E\left(\int\frac{dx}{S_1}-\frac{y}{b_2}\right)$, $z_2=\frac{1}{b_2} \int E(t)\,dt$

\item Spectral parameter:
 $\quad \lambda(y,t)=\lambda_0$

\item Reduced Fields:
\begin{eqnarray}&&   \psi(x,y,t)=\sqrt{\frac{S_1}{E}}\,
{\displaystyle e}^{{\displaystyle \left(\frac{a_0y}{b_2}\right)}}
{\displaystyle  e}^{{\displaystyle \left(\frac{a_0t}{b_2\lambda_0^n}\right)}}\,\Phi(z_1,z_2)\nonumber \\ && M(x,y,t)=\frac{E^2}{S_1^2}\,H(z_1,z_2)\nonumber\\
 && U^{[1]}(x,y,t)=\frac{S_1}{b_2}\,V^{[1]}(z_1,z_2)+S_1\frac{d}{dt}\left(\int\frac{dx}{S_1(x,t)}\right)+S_1\frac{E_t}{E^2}z_1\nonumber\\
&& U^{[j]}(x,y,t)= \frac{S_1}{b_2}\,V^{[j]}(z_1,z_2)\nonumber\end{eqnarray}

\item Reduced Spectral problem:
\begin{eqnarray}&&\Phi_{z_1z_1}+\left(\frac{\lambda_0}{2}H-\frac{1}{4}\right)\Phi=0\nonumber\\
&&\lambda_0^n\Phi_{z_2}=({\cal \hat B}-1)\Phi_{z_1}-\frac{{\cal \hat B}_{z_1}}{2}\Phi\label{5.7}
\end{eqnarray}
where
\begin{equation} {\cal \hat B}=\sum_{j=1}^n \lambda_0^{(n-j+1)}V^{[j]}(z_1,z_2)  \label{5.8}\end{equation}

\item Reduced Hierarchy: The compatibility condition of (\ref{5.7}) yields \textbf{the autonomous hierarchy}:
\begin{eqnarray}&&\frac{\partial^3 V^{[n]}}{\partial z_1^3}-\frac{\partial V^{[n]}}{\partial z_1}-\frac{\partial H}{\partial z_1}=0\nonumber\\
&&2H\frac{\partial V^{[1]}}{\partial z_1}+V^{[1]}\frac{\partial H}{\partial z_1}-\frac{\partial H}{\partial z_2}=0\\
&&2H\frac{\partial V^{[j+1]}}{\partial z_1}+V^{[j+ 1]}\frac{\partial H}{\partial z_1}+\frac{\partial^3
V^{[j]}}{\partial z_1^3}-\frac{\partial V^{[j]}}{\partial z_1}=0,\nonumber \label{5.9}
\end{eqnarray}
\end{itemize}
which is \textbf{the celebrated negative Camassa-Holm hierarchy \cite{AF88}, \cite{ch93}, \cite{chh94},
\cite{cgp97}}.
\subsection*{IV.3: $a_2\neq 0,  \quad E=\sqrt{A_1^2-4B_1C_1}=0$.}

\begin{itemize}
\item Reduced variables: $\quad z_1=\int\frac{dx}{S_1(x,t)}-\frac{\textrm{ln}(S_2)}{a_2}$, $z_2=t $

\item Spectral parameter:
 $\quad \lambda(y,t)=S_2^{\left(-\frac{1}{n}\right)}\Lambda(z_2)$
 where $\Lambda(z_2)$ satisfies the non-isospectral condition
 $$n\frac{d \Lambda(z_2)}{dz_2}+a_2\Lambda(z_2)^{(1-n)}=0$$

\item Reduced Fields:
\begin{eqnarray} && \psi(x,y,t)=\sqrt{S_1}\,\lambda^{{\displaystyle \left(-\frac{a_0n}{a_2}\right)}}\,\Phi(z_1,z_2)
\nonumber \\ && M(x,y,t)=\frac{S_2^{\left(\frac{1}{n}\right)}}{S_1^2}\, H(z_1,z_2)\nonumber\\
&& U^{[1]}(x,y,t)=S_1\,V^{[1]}(z_1,z_2)+S_1\frac{d}{dt}\left(\int\frac{dx}{S_1(x,t)}\right)\nonumber\\
&&U^{[j]}(x,y,t)= S_1\, S_2^{\left(\frac{1-j}{n}\right)}\,V^{[j]}(z_1,z_2)\nonumber\end{eqnarray}

\item Reduced Spectral problem:
\begin{eqnarray}&&\Phi_{z_1z_1}+\frac{\Lambda(z_2)}{2}H\Phi=0\nonumber\\
&&\Lambda(z_2)^n\Phi_{z_2}=({\cal \hat B}-1)\Phi_{z_1}-\frac{{\cal \hat B}_{z_1}}{2}\Phi\label{5.10}
\end{eqnarray}
where
\begin{equation} {\cal \hat B}=\sum_{j=1}^n \Lambda(z_2)^{(n-j+1)}V^{[j]}(z_1,z_2)  \label{5.11}\end{equation}

\item Reduced Hierarchy: The compatibility condition of (\ref{5.10}) yields \textbf{the autonomous hierarchy}:
\begin{eqnarray}&&\frac{\partial^3 V^{[n]}}{\partial z_1^3}-\frac{\partial H}{\partial z_1}+\frac{a_2}{n}H=0\nonumber\\
&&2H\frac{\partial V^{[1]}}{\partial z_1}+V^{[1]}\frac{\partial H}{\partial z_1}-\frac{\partial H}{\partial z_2}=0\\
&&2H\frac{\partial V^{[j+1]}}{\partial z_1}+V^{[j+ 1]}\frac{\partial H}{\partial z_1}+\frac{\partial^3
V^{[j]}}{\partial z_1^3}=0.\nonumber \label{5.12}
\end{eqnarray}
\end{itemize}
\subsection*{IV.4:  $a_2\neq 0, \quad E=\sqrt{A_1^2-4B_1C_1} \neq 0$.}

\begin{itemize}
\item Reduced variables: $\quad z_1=E\left(\int\frac{dx}{S_1(x,t)}-\frac{\textrm{ln}(S_2)}{a_2}\right)$, $z_2=\int E(t)\,dt $

\item Spectral parameter:
 $\quad \lambda(y,t)=S_2^{\left(-\frac{1}{n}\right)}\Lambda(z_2)$
 where $\Lambda(z_2)$ satisfies the non-isospectral condition
 $$n\frac{d \Lambda(z_2)}{dz_2}+\frac{a_2}{E(z_2)}\Lambda(z_2)^{(1-n)}=0$$

\item Reduced Fields:
\begin{eqnarray}  && \psi(x,y,t)=\sqrt{\frac{S_1}{E}}\,\lambda^{{\displaystyle\left(-\frac{a_0n}{a_2}\right)}}\,\Phi(z_1,z_2)\nonumber\\
&& M(x,y,t)=\frac{E^2}{S_1^2}\,S_2^{\left(\frac{1}{n}\right)}\,H(z_1,z_2)\nonumber\\
&&  U^{[1]}(x,y,t)=S_1\,V^{[1]}(z_1,z_2)+S_1\frac{d}{dt}\left(\int\frac{dx}{S_1(x,t)}\right)+S_1\,\frac{E_t}{E^2}\,z_1\nonumber \\
&& U^{[j]}(x,y,t)= S_1\,S_2^{\left(\frac{1-j}{n}\right)}\,V^{[j]}(z_1,z_2)\nonumber\end{eqnarray}

\item Reduced Spectral problem:
\begin{eqnarray}&&\Phi_{z_1z_1}+\left(\frac{\Lambda(z_2)}{2}H-\frac{1}{4}\right)\Phi=0\nonumber\\
&&\Lambda(z_2)^n\Phi_{z_2}=({\cal \hat B}-1)\Phi_{z_1}-\frac{{\cal \hat B}_{z_1}}{2}\Phi\label{5.13}
\end{eqnarray}
where
\begin{equation} {\cal \hat B}=\sum_{j=1}^n \Lambda(z_2)^{(n-j+1)}V^{[j]}(z_1,z_2)  \label{5.14}\end{equation}

\item Reduced Hierarchy: The compatibility condition of (\ref{5.13}) yields \textbf{the non-autonomous hierarchy}:
\begin{eqnarray}&&\frac{\partial^3 V^{[n]}}{\partial z_1^3}-\frac{\partial V^{[n]}}{\partial z_1}-\frac{\partial H}{\partial z_1}+\frac{a_2}{n E(z_2)}H+=0\nonumber\\
&&2H\frac{\partial V^{[1]}}{\partial z_1}+V^{[1]}\frac{\partial H}{\partial z_1}-\frac{\partial H}{\partial z_2}=0\\
&&2H\frac{\partial V^{[j+1]}}{\partial z_1}+V^{[j+ 1]}\frac{\partial H}{\partial z_1}+\frac{\partial^3
V^{[j]}}{\partial z_1^3}-\frac{\partial V^{[j]}}{\partial z_1}=0.\nonumber \label{5.15}
\end{eqnarray}
\end{itemize}
\section{Non classical symmetries for $\xi_3=\xi_2=0$, $\xi_1=1$}

We can  now write
\begin{eqnarray}
\psi_x&=& \phi_1\nonumber\\
 M_x&=& \Theta_0\label{6.1}\\
U^{[j]}_x &=&\Theta_j\quad j=1..n\nonumber
\end{eqnarray}
This is not a case of particular interest because the resulting symmetries are:
 \begin{eqnarray} \xi_1&=& 1\nonumber\\
 \xi_2&=& 0\nonumber\\
 \xi_3&=& 0\nonumber\\
  \phi_1&=& \frac{1}{2}\left(1\pm i\sqrt{2\lambda M}\right)\psi\label{6.2}\\
   \phi_2&=& 0\nonumber\\
 \Theta_0&=& -2M\nonumber\\
 \Theta_j&=& U^{[j]},\quad j=1..n,\nonumber
\end{eqnarray}
which holds only if $M_t=M_y=0$ and yields the following reductions:
\begin{eqnarray}
&& z_1=y,\,z_2=t\nonumber\\
&& \lambda(y,t)=\Lambda(z_1,z_2)\nonumber\\
&&\psi= {\displaystyle e}^{{\displaystyle \left(\frac{x\pm i\sqrt{2\Lambda H_0}e^{-x}}{2}\right)}}\Phi(z_1,z_2)\nonumber\\
&& M(x,y,t)=H_0\,e^{-2x}\label{6.3}\\
&& U^{[j]}(x,y,t)=e^x\,V^{[j]}(z_1,z_2),\quad j=1..n,\nonumber\\
\end{eqnarray}
and it is easy to see that (\ref{6.3}) satisfies (2.4) for $V^{[j]}(z_1,z_2),\, (j=1..n)$ arbitrary and $H_0$ constant.

\section{Conclusions}
\begin{itemize}
\item We  started with the spectral problem (although non-isospectral) associated with a Camassa-Holm hierarchy in $2+1$ dimensions
\item Non-classical Lie symmetries of this CHn2+1 spectral problem have been obtained.
\item Each Lie symmetry yields a reduced spectral $1+1$ problem whose compatibility condition provides a
$1+1$ hierarchy.
\item The main achievement of this paper is that our procedure also provides  the reduction of the eigenfunction as
well as the spectral parameter. In many cases, the reduced parameter also proves to be non-isospectral, even
in the $1+1$ reduction \cite{gp03}.
\item
There are several different reductions but they can be summarized in 9 different non-trivial cases: I.i
(i=1..5) and IV.j (j=1..4). Five of these hierarchies (I.3,I.4,I.5,IV.3,IV,4) have a non-isospectral Lax pair
and two of them (I.1 and IV.2) are the positive and negative Camassa-Holm hierarchies respectively. The
equations for all these reduced hierarchies have been explicitly written in each case.

\end{itemize}

\section*{Acknowledgements}
This research has been supported in part by the DGICYT under project  FIS2009-07880 and JCyL under contract
GR224.

\section*{Appendix} Let us go on to prove that the reduced hierarchies obtained by means of the
reductions related to the symmetries of type II are the same as type I, even though the reductions of
variables  and fields are different.
\subsection*{II.1) $A_1(t)=C_1(t)=0,\,B_1(t)\neq 0,\quad a_2=0,\, a_3=0,\, b_2= 0.$}
The reductions are now
\begin{itemize}
\item Reduced variables: $\quad z_1=-\textrm{ln}\left(e^{-x}+\frac{1}{b_3}\int B_1(t)\,dt\right) $, $\quad z_2=y $
\item Spectral parameter: $\quad \lambda(y,t)=\lambda_0$
\item Reduced Fields:
\begin{eqnarray} &&\psi(x,y,t)={\displaystyle e}^{{\displaystyle  \left(\frac{a_0t}{b_3}\right)}}\,{\displaystyle e}^
{{\displaystyle\left( \frac{\lambda_0^n
a_0z_2}{b_3}\right)}}\,\left(\frac{e^x}{e^{z_1}}\right)^{\frac{1}{2}}\,\Phi(z_1,z_2)\nonumber\\
&&M(x,y,t)=\left(\frac{e^{z_1}}{e^x}\right)^2\,H(z_1,z_2)\nonumber\\
 &&U^{[1]}(x,y,t)=\left(\frac{e^x}{e^{z_1}}\right)\,V^{[1]}(z_1,z_2)-\frac{B_1}{b_3}e^x\nonumber\\
 &&U^{[j]}(x,y,t)=\left(\frac{e^x}{e^{z_1}}\right)\,V^{[j]}(z_1,z_2),\nonumber\end{eqnarray}
 \end{itemize}
which yield the same spectral problem as in case I.1.

\subsection*{II.2) $A_1(t)=C_1(t)=0,\,B_1(t)\neq 0,\quad a_2=0,\, a_3=0,\, b_2\neq 0.$}
This case affords the reductions
\begin{itemize}
\item Reduced variables: $\quad z_1=-\textrm{ln}\left(e^{-x}+\frac{1}{b_3}\int B_1(t)\,dt\right) $, $\quad z_2=\frac{y}{b_2}-\frac{t}{b_3} $
\item Spectral parameter: $\quad \lambda(y,t)=\left(\frac{b_3}{b_2}\right)^{\left(\frac{1}{n}\right)}\lambda_0$
\item Reduced Fields:  \begin{eqnarray} && \psi(x,y,t)=
{\displaystyle e}^{\left({\displaystyle \frac{a_0t}{b_3}}\right)}\,{\displaystyle e}
^{{\displaystyle\left(\frac{\lambda_0^n a_0z_2}{1+\lambda_0^n
}\right)}}{\displaystyle\left(\frac{e^x}{e^{z_1}}\right)^{\frac{1}{2}}}\,\Phi(z_1,z_2)\nonumber \\
&& M(x,y,t)=\left(\frac{b_2}{b_3}\right)^{\left(\frac{1}{n}\right)}\,\left(\frac{e^{z_1}}{e^x}\right)^2\,H(z_1,z_2)\nonumber\\
&& U^{[1]}(x,y,t)=\left(\frac{1}{b_3}\right)\,\left(\frac{e^x}{e^{z_1}}\right)\,V^{[1]}(z_1,z_2)-\frac{B_1}{b_3}e^x\nonumber\\
&& U^{[j]}(x,y,t)=\left(\frac{1}{b_3}\right)\left(\frac{b_2}{b_3}\right)^{\left(
\frac{1-j}{n}\right)}\,\left(\frac{e^x}{e^{z_1}}\right)\,V^{[j]}(z_1,z_2),\nonumber\end{eqnarray}
 \end{itemize}
and the spectral problem is the same as in I.2.

\subsection*{II.3) $A_1(t)=C_1(t)=0,\,B_1(t)\neq 0,\quad a_2=0,\, a_3\neq 0,\, b_2=0.$}
In this case, the reductions are
\begin{itemize}
\item Reduced variables: $\quad z_1=-\textrm{ln}\left(e^{-x}+\int \frac{B_1(t)}{S_3}\,dt\right)$, $\quad z_2=a_3y $
\item Spectral parameter: In this case, the reduction of the spectral parameter is a non-trivial one that yields
$$\quad \lambda(y,t)=S_3^{\left(\frac{1}{n}\right)}\Lambda(z_2)$$

\item Reduced Fields:
\begin{eqnarray} &&  \psi(x,y,t)=\Lambda(z_2)^{{\displaystyle \left(\frac{a_0 n}{a_3}\right)}}
S_3^{{\displaystyle\left(\frac{a_0 }{a_3}\right)}}{\displaystyle\left(\frac{e^x}{e^{z_1}}\right)^{\frac{1}{2}}}\,\Phi(z_1,z_2)\nonumber\\
&& M(x,y,t)=S_3^{\left(-\frac{1 }{n}\right)}\,\left(\frac{e^{z_1}}{e^x}\right)^2\,H(z_1,z_2)\nonumber\\
 && U^{[1]}(x,y,t)=\frac{a_3}{S_3}\,\left(\frac{e^x}{e^{z_1}}\right)\,V^{[1]}(z_1,z_2)-\frac{B_1}{S_3}e^x\nonumber\\
&& U^{[j]}(x,y,t)=a_3\,S_3^{\left(\frac{j-1
}{n}-1\right)}\,\left(\frac{e^x}{e^{z_1}}\right)\,V^{[j]}(z_1,z_2),\nonumber\end{eqnarray}

\end{itemize}
and the spectral problem is exactly the same as in I.3.

\subsection*{II.4) $A_1(t)=C_1(t)=0,\,B_1(t)\neq 0,\quad a_2=0,\, a_3\neq 0,\, b_2\neq 0.$}
We have the following results
\begin{itemize}
\item Reduced variables: $\quad z_1=-\textrm{ln}\left(e^{-x}+\int \frac{B_1(t)}{S_3(t)}\,dt\right)$, $\quad z_2=\frac{a_3y}{b_2} -\textrm{ln}(S_3)$
\item Spectral parameter: In this case the reduction of the spectral parameter  yields
 $$\quad \lambda(y,t)=\left(\frac{S_3}{b_2}\right)^{\left(\frac{1}{n}\right)}\Lambda(z_2)
$$
 where $\Lambda(z_2)$ is the reduced spectral parameter:

    \item Reduced Fields:
\begin{eqnarray} && \psi(x,y,t)=\Lambda(z_2)^{{\displaystyle\left(\frac{na_0}{a_3}\right)}}\,S_3^{{\displaystyle \left(\frac{a_0}{a_3}\right)}}\,{\displaystyle\left(\frac{e^x}{e^{z_1}}\right)^{\frac{1}{2}}}\,\Phi(z_1,z_2)\nonumber \\
&&M(x,y,t)=\left(\frac{b_2}{S_3}\right)^{\left(\frac{1}{n}\right)}\,\left(\frac{e^{z_1}}{e^x}\right)^2\,H(z_1,z_2)\nonumber\\
&& U^{[1]}(x,y,t)=\frac{a_3}{S_3}\,\left(\frac{e^x}{e^{z_1}}\right)\,V^{[1]}(z_1,z_2)-\frac{B_1}{S_3}e^x \nonumber \\
 &&U^{[j]}(x,y,t)=\frac{a_3}{S_3}\left(\frac{S_3}{b_2}\right)^{\left(\frac{j-1}{n}\right)}\,\left(\frac{e^x}{e^{z_1}}\right)\,
 V^{[j]}(z_1,z_2),\nonumber\end{eqnarray}

\end{itemize}
which yields the spectral problem I.4.

\subsection*{II.5) $A_1(t)=C_1(t)=0,\,B_1(t)\neq 0,\quad a_2\neq 0$}
The same spectral problem I.5 is obtained through the following reductions
\begin{itemize}
\item Reduced variables: $\quad z_1=-\textrm{ln}\left(e^{-x}+\int \frac{B_1(t)}{S_3(t)}\,dt\right)$, $\quad z_2=S_2\,S_3^{{\displaystyle \left(-\frac{a_2}{a_3}\right)}}$
\item Spectral parameter: In this case the reduction of the spectral parameter  yields
 $$\quad \lambda(y,t)=S_3^{{\displaystyle \left(\frac{a_3-a_2}{a_3n}\right)}}\Lambda(z_2),$$
 where $\Lambda(z_2)$ is the reduced spectral parameter:

\item Reduced Fields:
\begin{eqnarray} &&  \psi(x,y,t)=\Lambda(z_2)^{{\displaystyle \left(\frac{na_0}{a_3-a_2}\right)}}S_3^{{\displaystyle \left(\frac{a_0}{a_3}\right)}}\,{\displaystyle\left(\frac{e^x}{e^{z_1}}\right)^{\frac{1}{2}}}\,\Phi(z_1,z_2)\nonumber \\
&& M(x,y,t)=S_3^{{\displaystyle \left(\frac{a_2-a_3}{a_3n}\right)}}\,\left(\frac{e^{z_1}}{e^x}\right)^2\,H(z_1,z_2)\nonumber\\
 &&U^{[1]}(x,y,t)=\left(\frac{a_2}{S_3}\right)\,\left(\frac{e^x}{e^{z_1}}\right)\,V^{[1]}(z_1,z_2)-\frac{B_1}{S_3}e^x\nonumber\\
&& U^{[j]}(x,y,t)=\left(\frac{a_2}{S_3}\right)S_3^{{\displaystyle
\left(\frac{(a_3-a_2)(j-1)}{a_3n}\right)}}\,\left(\frac{e^x}{e^{z_1}}\right)\,
 V^{[j]}(z_1,z_2).\nonumber\end{eqnarray}

\end{itemize}
\end{document}